# Development of a Modes of Collaboration framework

Alanna Pawlak[1], Paul W. Irving[1, 2], and Marcos D. Caballero[1, 2]
[1]Michigan State University, Department of Physics and Astronomy,
Biomedical and Physical Sciences, East Lansing, MI, 48824
[2]Michigan State University, CREATE for STEM Institute
620 Farm Lane, East Lansing, MI, 48824

**Abstract**. Group work is becoming increasingly common in introductory physics classrooms. Understanding how students engage in these group learning environments is important for designing and facilitating productive learning opportunities for students. We conducted a study in which we collected video of groups of students working on conceptual electricity and magnetism problems in an introductory physics course. In this setting, students needed to negotiate a common understanding and coordinate group decisions in order to complete the activity successfully. We observed students interacting in several distinct ways while solving these problems. Analysis of these observations focused on identifying the different ways students interacted and articulating what defines and distinguishes them, resulting in the development of the Modes of Collaboration framework. The Modes of Collaboration framework defines student interactions along three dimensions: social, discursive, and disciplinary content. This multi-dimensional approach offers a unique lens through which to consider group work and provides a flexibility that could allow the framework to be adapted for a variety of contexts. We present the framework and several examples of its application here.

## INTRODUCTION

Interactive instruction is becoming increasingly common in introductory physics classes, with more instructors implementing techniques such as small group discussions, group problem solving, and team-based projects. These techniques have been found to be more effective by some metrics [1-5], but a number of aspects remain ill understood [6,7]. Much work remains to be completed to better understand these interactive learning environments and the effects they have on student learning [8-10].

Several lenses aimed at understanding various aspects of how students engage in such work have been developed. Some seek to assess the impact of collaborative work on individual student learning [1,5,7,11]. Others attempt to understand the different ways students may perceive group work [12-15]. Still others aim to identify ways to optimize group work [16-18]. Another broad area of investigation has endeavored to categorize the ways that students engage with group work. Such work has approached this goal in a variety of ways, including a particular focus on social aspects of group work [4,19,20], discursive aspects of group work [18,21,22,23], and framing aspects of group work [10,11,24,25]. In this paper, we present a new framework called the Modes of Collaboration that attends to three dimensions: social, discursive, and disciplinary content. We did not make use of framing as a dimension directly, as we found that considering the ways in which students discussed physics content was better able to capture our observations, however, this dimension does have a relationship to framing (see Framework section).

By attending to all three of these dimensions simultaneously and independently, the Modes of Collaboration is a framework that is simple to apply, but that still provides multi-faceted insight into students' engagement with group work. In addition to presenting the framework here, we identify four specific Modes that students engaged in within our context based on observation of a small set of video data from one day of class work. The Modes of Collaboration framework presented here acts as a proof that student group work can be described along the three dimensions, and we propose that it is flexible enough to be used in other contexts beyond the one analyzed here.

## BACKGROUND

Group work has long been an area of investigation in physics education and education more generally, and has been studied from many different perspectives. Some work focuses on individual content understanding, typically using pre and post measures of individual learning to measure the impact of group work [1,5,7,11]. Other studies have attended to student perceptions of group work, usually through the observation of student behavior in groups or through interviews with students, aiming to more qualitatively understand students' experiences [12-15]. Research has also been conducted on ways that group work might be optimized to achieve the best outcomes for students, for example, considering which types of activities or which group compositions result in the greatest learning gains for students [16-18]. Finally, a great deal of work has been done with the aim of developing ways to categorize student engagement with group work. This has been done in a variety of ways, including attending primarily to social dynamics [4,19,20], student discourse [18,21,22,23], or student perceptions and framing [10,11,24,25]. Below, we offer additional background on the work conducted along each of these lines to frame and situate the Modes of Collaboration framework. While much of this previous work is not specific to undergraduate students or to physics, it is nonetheless valuable in informing and providing context for our work. Elementary school students and college students are of course different in many ways, and even college math students and college physics students are different, but themes found in one of these populations can still be productive when considering the other. We describe some of these themes below, and demonstrate their utility in our context in the Framework and Modes of Collaboration sections.

**Individual Content Understanding**

Much early research surrounding group work focused on identifying or assessing the impact of collaborative work on individual students' understanding of the content. In a study of high school science students, Amigues compared the individual post-test performance of students who worked on a preceding activity alone to students who worked on the preceding activity in dyads, and found that the students who worked in dyads were more successful on the post-test [1]. In a later study specific to college physics students, Heller et al. tracked how students' individual problem solving abilities developed over the duration of a semester-long course that implemented collaborative group problem solving sessions, and found that their individual problem solving abilities improved [3].

In efforts to similarly study the impact that group work has on individual student outcomes, many studies have made use of concept inventories as a metric. These concept inventories [26-28] are multiple-choice exams centered on a particular content area, which students complete individually. While they are unable to capture many aspects of a student's experience, they have historically been used as an indicator of a student's understanding of the concepts probed. Lumpe et al. used the Photosynthesis Concept Test (PCT) to compare the learning outcomes of high school students who completed a task on photosynthesis individually to students who completed the same task in a group. They found that students who had worked in groups were more successful on the PCT, but that not every member of a given group experienced the same degree of improvement [4].

Studies such as these have been important in providing evidence of group work's efficacy, albeit as measured by metrics that have limitations [29-31]. What they cannot provide, however, is an understanding of what happens *during* group work. By focusing on comparisons between students' pre- and post- instruction understanding, they neglect to consider the ways in which students interact and speak while engaging in group work; the social and discursive dimensions. Furthermore, a strict focus on individual content understanding does not account for the aspects of learning that may occur beyond purely cognitive models [32].

**Student Perceptions**

In addition to identifying the ways in which group work impacts individual student content understanding, efforts have been made to examine how students perceive group work. In a study of elementary students working on a science task in groups, Anderson et al. conducted observations and interviews to understand students' goals and feelings of success or lack thereof during the activity [12]. They defined three areas on which students may focus their attention when working in groups: task structure and accountability systems, interpersonal relationships, and scientific activity. Based on Anderson et al.'s observations and interviews with students after the activity, they found that the students focused primarily on interpersonal relationships and task structure goals, and that students did not appear to focus as much on the scientific activity component. Additionally, Anderson et al. found that students felt that they were successful at the activity overall. Grindstaff et al. also examined students' perceptions of collaborative work by interviewing students working on short-term research projects with peers [15]. They found that students discussed several types of support that peers may provide when working collaboratively: emotional, technical, and cognitive, and that there was a great variance in which ones students reported as being most relevant or important.

Student perceptions of group work have also been examined through the construct of epistemological framing [13,14,24,25,34,35]. Scherr et al. developed this construct by building upon previous work on framing in general, which they characterized as "how an individual or group forms a sense of 'What is it that's going on here?'" From this, they narrowed to examining how students frame activities specifically related to knowledge, and called this epistemological framing [14]. In analyzing the discussion of undergraduate physics students working on tutorials, they observed four behavioral clusters that they then mapped to different epistemological framings. These frames were: discussion, worksheet, TA, and joking. In another study of students working collaboratively, Irving et al. proposed an alternative way to understand students' epistemological framing [13]. In their work, they define two axes with which to characterize students' framing, rather than the discrete categories presented by Scherr et al. One axis describes the degree to which a student's statements are serious or silly, and the other describes the degree to which a student's discussion is narrow versus expansive.

This type of work on students' perceptions of the tasks and goals in group work and their perceptions of group work itself provides a complementary perspective to the insights gained by studies of individual student content understanding. While the latter attends to outcomes, the former attends to the process and experiences of students. Studies of student perceptions of group work frequently focus on social aspects, sometimes discuss discursive aspects, and less frequently consider disciplinary content-related aspects of group work, but rarely attend to all three simultaneously.

**Optimizing Group Work**

There have also been efforts made to identify the ways to best design and implement group work to maximize its benefit to students. Heller et al. examined the effect that group makeup had on the success of groups [16]. Their results indicated that heterogeneity with respect to incoming ability produced the highest rate of success for groups. They also found that groups that were homogenous with respect to gender, or groups in which there were more women than men were most successful. In addition to examining group makeup, they investigated the impact of giving students explicit guidance on how to work in groups. Students were given one of three roles: manager, skeptic, and checker/recorder. They found that assigning these roles reduced issues of individuals becoming too dominant in groups, or groups being conflict-avoidant. They also found that giving students time to engage in explicit discussion of their group's interactions was beneficial. Van Boxtel et al. examined the impact that having students complete individual preparatory work before working in groups had on their success in those groups [18]. Their results indicated that the individual preparatory work led to students asking each other more questions while working together, and improved individual learning gains measured after the group work. Webb et al. sought to identify the conditions that must be met in order for students working in peer-directed groups to give and receive help in a productive way [17]. They propose that in order for received help to be effective it must be relevant, timely, correct, and sufficiently elaborated. They also identify three more conditions necessary for received help to be effective: the recipient must understand the help, the recipient must have a chance to make use of the help, and the recipient must act on that chance to use the help. The authors go on to use these conditions to identify the responsibilities of the help-seeker, help-giver, and teacher in making these conditions possible.

Studies seeking to optimize group work are essential to providing students with the best learning experiences in collaborative environments. Their results can directly inform instructional choices. Similar to the studies on individual content understanding, however, they frequently compare students' pre- and post- performance, and do not attend to the processes that occur during group work. In doing so, they typically do not attend to the discursive and social dimensions of group work.

### Categorizing Student Engagement with Group Work

Another area of investigation has endeavored to categorize the various ways that students engage with group work. Some of this work has focused primarily on the social dimension when developing categories. In the work by Lumpe et al. previously discussed, they identified two general interaction styles that students may experience when collaborating: consonant, or generally agreeable, and dissonant, or generally negative [4]. Roth et al. also attended to social factors, and categorized the ways in which students may navigate a disagreement, observing that they would proceed in several distinct ways: collaborative construction, adversarial exchanges, and the formation of temporary alliances [20]. Richmond et al. identified the different social roles that students took on when working in groups. They observed four social roles: leader, helper, active non-contributor, and passive non-contributor [19]. They further found that students taking on a leadership role would lead in one of three styles: inclusive, persuasive, or alienating.

Others seeking to categorize students' engagement in group work have done so by attending to the discursive dimension. Hogan et al. examined the ways that students speak to one another when working in peer guided group discussions and teacher guided group discussions [22]. They identified four modes of discussion: peer knowledge construction, teacher guided knowledge construction, logistical conversation, and off task conversation. They found that the relative occurrence of these modes varied greatly across groups. One hypothesis they offered for this result was the teacher spending more or less time with a group depending on their level of prior knowledge. In the Van Boxtel work described earlier, the authors identified several styles of student interactions based on their discourse: question, conflict, and reasoning [18]. They found that questioning episodes occurred most frequently, and that reasoning episodes were most likely to lead to elaboration of ideas. Haussman et al. also made use a categorization scheme attending to discourse [21]. They analyzed three proposed mechanisms of collaborative learning: other directed explaining, self directed explaining, and co-construction. The authors found that co-construction led to high individual learning gains for both participants as measured by an individual post test, and self directed and other directed explaining led to higher individual learning gains for the students giving the explanations. Other directed explaining was more effective for the listener than self directed explaining was for the listener.

Students' engagement with group work has also been categorized by attending to their perceptions of collaboration and the activity on which they are working. As described earlier, both the work of Scherr et al. and Irving et al. set forward frameworks with which to classify students' behavior in groups based on their epistemological framing [13,14].

All of these methods of categorization provide valuable frameworks to understand students' engagement with group work. Applying such categories can offer a way to make sense of what students find important, how they perceive each other, and how they speak to one another. As outlined though, these frameworks typically attend to only one dimension of group work. A framework that categorizes students' engagement with group work while simultaneously attending to discursive, social, and disciplinary content dimensions can provide insights that focusing on only one may be unable to provide. The Modes of Collaboration framework attempts to do just this, outlining several distinct Modes of student interaction that are characterized by their discourse, their social interactions, and their engagement with the disciplinary content.

## STUDY CONTEXT

The data presented in this paper were collected from an introductory electricity and magnetism course at a large university. There were approximately 120 students in the course, and most were sophomore life-science majors. The students all attended lecture three times a week, and a laboratory session once a week in sections of approximately 20 students. In these laboratory sessions, students worked collaboratively in groups of three or four on a variety of activities depending on the week (traditional labs, tutorials, conceptual and calculational problems, etc.). For this study, nine unique groups of students were video recorded as they worked during their weekly laboratory sections, resulting in a total of 2 hours and 29 minutes of video.

The activity they are working on is a series of three conceptual questions about the electric field and electric potential energy in the area around different distributions of charge. Students would first read the problem statement without knowledge of the charge distribution in question. They would then select as a group a representation of the charge distribution (electric field lines, electric field vectors, or electric potential lines) to view, based on what they thought would be most helpful in answering the problem. An instructor would then bring them their requested representation, and the students would use it in order to answer the problem. Each question had a well-defined correct answer, but could be solved using multiple methods. For example, one might use the representation to deduce where the source charges are located, then use this information to answer the question, or one might use the representation directly to answer the question, without considering the location of the source charges. This activity was similar to those they had experienced in previous laboratory sessions in that it required cooperative group work on conceptual problems, but was unique in that it required a level of explicit planning and strategy (in the selection of a representation) that was not typically necessary.

## METHODS

Analysis began with multiple coarse viewings of the video data, attending to instances of explicit interaction among students where their discussion related to the activity. An explicit interaction is one that involves two or more students speaking, and an interaction that is related to the activity is one wherein the content of the students' speech was related to the physics content present in the activity. Focusing on these segments reduced our data to 1 hour and 11 minutes of video. Once these segments were identified, they were viewed successively, seeking emergent trends in the student interactions and behavior observed. In these emergent trends, distinct patterns were found that related to the three dimensions: social, discursive, and disciplinary content. The data was then split into "episodes", which were defined by a shift in a group's behavior along any of the three dimensions. For example, in a given segment of video, if a group's interaction with the disciplinary content appeared to change, this would be considered the end of one episode, and the beginning of a new episode. For each episode, the students' interactions were characterized along each dimension independently, and then episodes having all three dimensions in common were grouped and termed Modes. Preliminary definitions for each Mode were crafted based on exemplars, then refined through multiple viewings of every illustrative episode.

Similar to the work of Hogan et al., our procedure of analysis was not predetermined, instead emerging from our observations, and quantitative inter-rater reliability in the identification of Modes or dimensions was not our goal [22]. Our goal was to craft definitions for the Modes emerging from this data set that qualitatively described the speech and behavior seen in each Mode. The first author's knowledge of the data made her best suited to making such identification and analysis decisions, and the co-authors provided critical feedback on these decisions throughout the development of the framework. Through iterating on this process, we arrived at robust Mode definitions that were able to capture the commonalities seen in each instance of each

Mode, and also accounted for the differences seen among Modes.

## THE FRAMEWORK

The Modes of Collaboration are defined along three dimensions: social, discursive, and disciplinary content. Each individual Mode is characterized in a particular way within each of these dimensions.

### Social Dimension

The social dimension addresses the overall tenor of the students' interactions with one another. It accounts for the atmosphere in which the students' conversation takes place. To describe the social dimension, we make use of the interaction styles identified by Lumpe and Staver [4]. In their work, they observed that groups of students would interact in ways that were consonant or dissonant. Consonant interactions were characterized by agreement among students and the validation of peers' ideas, while dissonant interactions were characterized by conflict (taken here to mean explicit disagreement), a lack of recognition, or criticism of peers' ideas. It is important to note that consonant interactions are not necessarily better than dissonant interactions. For example, respectful critiques of one another's reasoning can lead students to a more robust understanding. Furthermore, in selecting this scheme of categorization for the social dimension, we do not claim that it fully describes the richness of students' social interactions. Rather, the assessment of consonant versus dissonant provides a simple and productive way to determine the general tone of a group's discussion, which is what we define the social dimension of the Modes of Collaboration framework to be.

### Discursive Dimension

The discursive dimension deals with the way in which students communicate with each other. It describes the ways that students present their ideas and the structure of their conversation. The discursive dimension is grounded in Hogan's work on knowledge construction and Toulmin's work on argumentation [22,33]. In their work, Hogan et al. identified three modes that described the interactions of their students: knowledge construction (peer or teacher-guided), logistical, and off task. Knowledge construction refers to when the conversation is related to scientific phenomena, logistical refers to when students discuss tasks necessary to complete the activity but not related to scientific content, and off task refers to when students discuss things unrelated to the task entirely. Since we sought to develop a framework that describes students' interactions when they are discussing physics, the Modes of Collaboration all occur within the knowledge construction mode identified by Hogan et al. We therefore could not use their three modes as a scheme with which to further analyze the student interactions we were interested in, but we did make use of an element of their analysis. In developing their three modes, Hogan et al. identified three interaction patterns. These interaction patterns were consensual, responsive, and elaborative.

Consensual interactions are those in which only one student makes substantive contributions, while other students simply agree, accept (explicitly or passively), or repeat the contributions of that student. Responsive interactions are those in which multiple students make substantive statements. Elaborative interactions are those in which multiple students make substantive statements, and those statements build off preceding statements by making connections between ideas, correcting someone's idea, or disagreeing with someone's idea and providing a counterargument. In addition to Hogan et al.'s interaction patterns, we used argumentation as a way to characterize student discourse. As conceptualized by Toulmin, argumentation is composed of evidence, a claim made based on that evidence, and warrants justifying how the evidence supports the claim. If students were observed to make use of these elements when presenting their ideas, their discourse was characterized as argumentation. It is worth noting that argumentation falls under the elaborative interaction pattern described by Hogan et al. Nonetheless, the choice to supplement Hogan et al.'s interaction patterns with Toulmin's argumentation was made because separating discourse that had formal argumentation from that which had a "non-argumentative" elaborative interaction pattern was productive. The distinction allows the Modes of Collaboration framework to attend to the difference between a series of unsubstantiated ideas shared by students, and a sequence of explicitly supported claims. These two interactions may suggest different motives for the students and may have different results for their further interactions, and we therefore wanted to capture this difference in the discursive dimension.

**Disciplinary Content Dimension**

The disciplinary content dimension addresses the ways that the students discuss physics. It describes the types of physics content on which their conversations focus. In the previously mentioned work by Irving et al. on epistemological framing, they identified an axis that described the scope of students' framing, with narrow at one end and expansive at the other end [13]. We do not make direct use of this construct, as our work does not attempt to identify the ways in which students frame the activity. Instead, we define students' discussion to be related to *specific* physics content or *abstract* physics content. These terms were selected in order to convey the degree to which the content being referenced is tied concretely to the situation the students are analyzing. When students discuss specific physics content, they focus on physics content applied specifically to the question at hand, such as discussing the orientation of the electric field vectors in the diagram presented in the question. Abstract physics content discussion refers to discussion that is not directly related to producing an answer to the activity question. Instead, it centers on concepts in general, such as a discussion of the meaning of electric potential energy. Thus, based on the scope of the physics content that the students discussed, the disciplinary content of their discussion was characterized as either abstract or specific.

A Mode is defined by its classifications in each of the three dimensions. It is worth noting that the dimensions are treated independently. Each dimension is assessed solely based on the discourse and behaviors observed, without reference to categorizations made along the other dimensions. It could be that some combinations do not occur (consonant social dimension and argumentation discursive dimension appear contradictory, for example), but by coding across each dimension individually, we do not make any assumptions about such connections. It is also important to recognize the grain size that the Modes of Collaboration framework considers. When applying the framework, data is broken into episodes defined by apparent shifts in a group's interaction along any of the three dimensions. It is these demarcated episodes that are then analyzed along the three dimensions. A summary of the dimensions and the categorizations possible within each appears in the table below.

| Dimension | Social | Discursive | Disciplinary Content |
|---|---|---|---|
| Categorizations | Consonant, Dissonant | Consensual, Responsive, Elaborative, Argumentation | Specific, Abstract |

TABLE I. Summary of the Three Dimensions

**THE MODES OF COLLABORATION**

In this study, we identified four distinct ways in which students interacted: Debate, Informing, Co-construction of an Answer, and Building Understanding Towards an Answer, which are described below in detail. Table II provides an overview. The following subsections each begin with a description of how the Mode manifests based on student behavior, then outlines how it is defined using the three dimensions.

**Debate**

In the Debate Mode, two students engage in a dialogue, arguing their conflicting understandings of the concepts or responses to the activity prompts, while the remaining students in the group do not speak. It will continue until one student ultimately capitulates and accepts the other's reasoning, or at least ceases to argue their own. The Debate Mode is characterized by dissonant interactions in the social dimension, argumentation in the discursive dimension, and can be characterized by specific or abstract content in the disciplinary content dimension. The two episodes presented here happen in immediate succession, and demonstrate an example of specific Debate and an example of abstract Debate. In the episodes, Lindsay and Michael work on a problem that asks them to determine if there are any points of equilibrium in the area surrounding an electric quadrupole using an image of the electric field vectors. Immediately before the transcript begins, Lindsay has argued that the electric field vectors cancel.

Episode A1
1 Michael: Yeah, they're [electric field vectors]
2 not in opposite directions. They all are going
3 in the same direction, so they're not
4 cancelling.
5 Lindsay: Well no no no, they are. These are,
6 [gesturing at the electric field vectors on the
7 image]
8 because this one's going this way, and this
9 one's going this way. And then these two are
10 going in opposite directions, as well. So, they
11 are in opposite directions.

Episode A2
12 Michael: Well, you remember the tutorial
13 homework we did? And it had equal
14 magnitude charges, but one was negative
15 and one was positive, and they added
16 together.
17 Lindsay: Yeah… [no longer pointing at the
18 image]
19 Michael: Ones that are of the same sign and
20 equal magnitudes cancel out.
21 Lindsay: Why would they cancel if they're
22 the same sign, when you add them together?

Social Dimension
When engaged in the Debate Mode, the two active students interact in a dissonant way. For the duration of the Mode, the two Debating students explicitly disagree with one another. Rather than validating each others' ideas, they put forth criticism. In the examples, there are multiple instances of explicit disagreement. In lines 1-4, Michael's first response to Lindsay's argument that the field vectors cancel, he directly disputes her claim. Following this, in lines 5-11, Lindsay reiterates her belief, contradicting Michael's ideas. In addition to this overt conflict, Lindsay and Michael explicitly critique each other's ideas. In lines 1-4, Michael not only disagrees with Lindsay, but also explains the flaw he finds in her ideas. Similarly, in lines 5-11, Lindsay explains why she believes Michael is incorrect. Finally, in lines 21-22, Lindsay pushes back against a perceived weakness in Michael's reasoning. These critiques, along with the conflict present in these episodes, indicate that they both take place with a dissonant interaction style.

Discursive Dimension
Along the discursive dimension, the Debate Mode is characterized by argumentation. The two Debating students making use of evidence, claims, and warrants as they present their ideas to one another. In the examples, we see Michael and Lindsay make use of these elements of argumentation in their discussion. In lines 1-4, Michael presents both his evidence, the electric field vectors are pointing in the same direction, and his claim, that they do not cancel. Later, in lines 12-16, he provides his warrant. He refers to a previous assignment to provide further information on why vectors going in the same direction do not indicate cancelling. In lines 5-11, Lindsay presents her competing claim, that the vectors do cancel, and evidence, that the image shows the vectors pointing in opposite directions. Such implementation of claims, evidence, and warrants indicates that the students are engaging in argumentation in these episodes.

Disciplinary Content Dimension
The Debate Mode may be specific or abstract in the disciplinary content dimension. Students may discuss physics content directly related to the question at hand, or they may focus their conversation on general cases and abstract concepts. In the examples, we see both kinds of disciplinary content present. In lines 1-11, Michael and Lindsay both attempt to directly answer the question. Each refers to the vectors on the image representing the charge distribution the question requires them to consider. Thus they consider specific physics content. In lines 12-20, the focus of their discussion shifts. Michael no longer refers to the question in the activity. He expands the conversation to discuss charges and "cancelling" more generally. In this way, the conversation now focuses on abstract physics content, indicating the beginning of a new episode. The students' behavior has not shifted along the social or discursive dimensions, however, and so Episode A2 is still Debate, but now a Debate of abstract content.

**Informing**
In the Informing Mode, one student, the "Informer", explains his or her ideas about the question at hand to one or more other group members. The Informer consistently offers his or her thoughts while other students do not; instead they only listen to or ask questions of the Informer. This continues until all participating group members begin writing the results of the discussion on their worksheets. The Informing mode is characterized by consonant interactions in the social dimension, consensual interaction patterns in the discursive dimension, and specific content in the disciplinary content dimension. In the example presented here, Jim, Erin, and Angela engage in the Informing Mode, with Erin acting as the Informer, as they attempt to determine how to maximize electric potential energy when placing a test charge near a given charge distribution, using a diagram of the equipotential lines surrounding the distribution.

Episode B
1 Erin: So all those rings [pointing at
2 equipotential lines] show the same potential
3 energy – one right. So the one that is the
4 smallest ring has the most potential energy.
5 Angela: Ohh.
6 Jim: Okay, so we just put it [the test charge]
7 right in the middle? [looking at Erin]
8 Erin: On the smallest ring.
9 Jim: On the smallest ring.
10 Angela: On the smallest ring.
11 Erin: Yeah.
12 [All three begin quietly writing on their
13 worksheets.]

Social Dimension
In the Informing Mode, students interact in a consonant way. There is no explicit conflict among group members or criticism of peers' ideas. Instead, the ideas presented are recognized and validated without resistance. In the example, Erin presents her thoughts about the answer to the question in lines 1-4. After this statement, the conversation contains implicit validation of Erin's statement by Angela in line 5, and explicit validation of Erin's idea by both Jim and Angela in lines 9 and 10, respectively. There is no criticism or rejection of Erin's ideas at any point, and no other students present ideas that could be subject to criticism or rejection. Thus we see that this episode is characterized by a consonant interaction style.

Discursive Dimension
During the Informing Mode, students' discourse is characterized by consensual interactions. The Informer is the only student who makes substantive contributions to the conversation, while the other students either explicitly agree and accept the Informer's statements, or ask short clarifying questions. In the example, Erin is the only student who makes a substantive contribution with her statements about the equipotential lines in the diagram in lines 1-4. In contrast, Angela responds with an implicit acceptance of Erin's statement in line 5, and in line 10, a direct repetition of a statement made by Erin. Jim's contributions consist of a question in line 6 clarifying Erin's initial statement, and in line 9, a direct repetition of Erin's statement. In this way, we see that this episode has a consensual interaction pattern, with Erin acting as the substantive contributor.

Disciplinary Content Dimension
In the Informing Mode, the disciplinary content of the students' conversation is characterized by specific content. The students discuss physics content as it relates directly to producing an answer to the question at hand. They do not discuss physics concepts in the abstract or expand their conversation to general cases. In lines 1-4 of the example, Erin presents an answer to the question the group is discussing. She does not discuss the meaning of electric potential energy or the function of equipotential lines in general. The only other statement in the episode that is not a simple assent or repetition is Jim's question in line 6. With this question, he confirms the answer that Erin is proposing, still limiting the scope of the physics content being discussed to the question at hand, and not abstract concepts. Accordingly, we see that the conversation in this interaction centers on specific content.

**Co-construction of an Answer**
In the Co-construction of an Answer Mode, two or more students work towards creating an answer to the question on which they are working. As the students work towards this answer, nearly every contributed statement is acknowledged and built upon. A student is considered to be participating in the Co-construction of an Answer if he or she makes statements relevant to the conversation during the episode. Non-participating students may or may not appear to be paying attention, but are not considered a part of the Co-construction of an Answer regardless, as they are not aiding in the *construction* of the answer. The Co-construction of an Answer Mode is characterized by a consonant interaction style in the social dimension, elaborative interaction patterns in the discursive dimension, and specific physics content in the disciplinary content dimension. In the example here, all four students in the group engage in the Co-construction of an Answer Mode as they consider which way a test charge would move when placed near an electric dipole using an image of electric field vectors.

Episode C
1 Lindsay: So, if we put a negative charge here
2 at this X, it's asking you which way it would
3 move after it's released.
4 Michael: It would move towards the outside
5 middle.
6 Lindsay: What do you mean by outside
7 middle? [looking at Michael]
8 Michael: Well, it's moving… [gestures hand
9 over the image, then pulls back, hesitating]
10 Lindsay: [looking at Michael, then speaking]
11 This [the image] is the electric field at each
12 point…
13 George: Wouldn't this be, this [gesturing at a
14 field vector] is showing where a positive test
15 charge would go, so wouldn't the electric
16 charge move opposite?
17 Lindsay: Well it's a negative, so wouldn't it
18 just move…[takes the image and starts to 19 draw on it]
20 Oscar: This is negative though, right here
21 [indicating a point on the image], isn't it?
22 Lindsay: Oh, it is negative; you're right, ok
23 [erases what she's drawn] so it would
24 move… opposite.

Social Dimension

The Co-construction of an Answer Mode is characterized by consonant interactions in the social dimension. There is no explicit conflict in the group's discussion, and the ideas that group members put forth are acknowledged and validated. In the example, nearly every statement made is acknowledged by the following statement. For example, Michael proposes an answer in lines 4-5, and in her clarifying question in lines 6-7, Lindsay makes direct reference to Michael's statement. In lines 10-12, while Lindsay does not explicitly acknowledge Michael's attempt to answer her clarifying question, the fact that she waits for Michael to trail off, then begins her statement by looking at him indicates that she is attempting to aid him in his hesitation. In the episode, we also see the explicit validation of peer ideas. In line 22, Lindsay specifically says, "you're right" in response to the idea that Oscar has presented in lines 20-21. Beyond the recognition and validation of peer ideas, there is also no explicit conflict present in the group's discussion. At no point do any of the students outright disagree with something another student has said. The closest statement to a disagreement in the episode comes from Oscar in lines 20-21, where he points out to Lindsay that a point charge she had been considering positive is in fact negative. Even in this statement, however, Oscar does not present his correction as a disagreement. Instead, he simply offers a new idea, phrasing it as a question, and not a rejection of Lindsay's understanding. This lack of explicit conflict, and the recognition and validation of peers' ideas indicate that the interactions in this episode are consonant.

Discursive Dimension

The discursive dimension of the Co-construction of an Answer Mode is characterized by elaborative interaction patterns. All participating students not only make substantive contributions to the discussion, but also explicitly connect those contributions to those of the other students. In the example, all four students in the group provide statements that are relevant to their discussion of the question and also explicitly relate their statements to each other's. In lines 10-11, Lindsay provides her understanding of what the image is showing them, thus offering a substantive contribution to the conversation. In lines 13-15, George responds by building off of this, offering a more specific understanding of what the electric field vectors show. With this response, he contributes substantively to the conversation, and also explicitly connects his ideas to what Lindsay has contributed. Earlier in the conversation, Michael provides a substantive contribution in lines 4-5 when he proposes an answer to the question, and in lines 8-9, provides another substantive contribution that is directly connected to the question Lindsay asks him in lines 6-7. In lines 20-21, Oscar also contributes a substantive and explicitly connected statement when he offers his correction of Lindsay's thoughts in lines 17-19. These substantive contributions and the explicit connections that the students make among them are what show this to be an elaborative interaction pattern.

### Disciplinary Content Dimension

The disciplinary content dimension of the Co-construction of an Answer Mode is described by specific physics content. The students focus their conversation on physics as it relates directly to the question they are working on, and do not attempt to expand their conversation to general cases or abstract concepts. In the example, the whole of the discussion is centered on producing an answer to the question, "which way will the charge move?". In lines 4-5, Michael provides a possible answer. In lines 10-15, Lindsay and George discuss the image of the charge distribution that the question asks them to consider, with Lindsay describing what the image shows, and George describing what the individual vectors on the image show. In neither case do they make claims about what electric fields or field vectors show in general, instead they refer specifically to the image they have. In lines 15-19, Lindsay and George both use the image to propose answers to the question. In response to Lindsay's answer, Oscar brings the group's attention to an element of the image in lines 20-21. Finally, Lindsay incorporates this and presents an answer again in lines 22-24. Throughout this episode, the students' conversation is focused on producing an answer to the question using information from the image they have been provided. This attention to the question and the absence of discussion of the meaning of physics concepts in the abstract indicate that this conversation is characterized by specific physics content.

## Building Understanding towards an Answer

In the Building Understanding towards an Answer Mode, two or more students discuss physics concepts in a way that is not directly related to answering a component of the activity, instead seeking to develop an understanding of the underlying concepts. Due to the nature of the activity the students in our data completed, this understanding was ultimately be aimed at answering a question in the activity, but nonetheless, the Building Understanding towards an Answer Mode focuses first on developing an understanding. Similar to Co-construction of an Answer, during Building Understanding towards an Answer, nearly every statement contributed is recognized and built upon. Also as in the case of Co-construction of an Answer, a student is only considered to be participating in the Building Understanding towards an Answer Mode if they verbally contribute to the discussion, as they otherwise are not contributing to the building of the group's understanding. The Building Understanding towards an Answer Mode is characterized by consonant interactions along the social dimension, elaborative interaction patterns along the discursive dimension, and abstract physics content along the disciplinary content dimension. In the example here, Leslie, Ben, and Ron engage in the Building Understanding towards an Answer Mode as they discuss the meaning of electric potential energy and its relationship to electric field lines while they work on a question asking them to determine how to maximize electric potential energy when placing a test charge near a given charge distribution, using a diagram of the electric field lines surrounding the distribution.

### Episode D
1 Leslie: What does it mean though to have
2 electric potential energy?
3 Ben: Remember here we did that question?
4 [flipping to a previous page in the activity] It
5 was a question where you compared the
6 electric potential energy between like A and
7 B, and the answer was A, here, has the
8 greater electric potential energy. So I think
9 it's how close you are to the actual…
10 Ron: [looking at Leslie] In other words, how
11 much energy you need to put in to like move
12 it.

### Social Dimension
The social dimension of the Building Understanding towards an Answer Mode is described by consonant interactions. The participating students recognize and validate one another's ideas, and there is no explicit conflict present in the discussion. In the example, Leslie begins the conversation by asking the group a question in lines 1-2. Ben acknowledges Leslie's question and engages with it when he answers her in lines 3-9. After this, in lines 10-12, Ron acknowledges both Leslie's question and Ben's answer by providing another answer to Leslie's question, and by framing it as "in other words" to Ben's answer. In addition to this consistent recognition of each other's statements, the episode also demonstrates a lack of conflict. In lines 3-12, neither Ben nor Ron disagree with one another's answers to Leslie's question. This lack of conflict and the students' acknowledgement of their peers' contributions show this to be a consonant interaction style.

### Discursive Dimension

The discursive dimension of the Building Understanding towards an Answer Mode is characterized by elaborative interaction patterns. All students participating contribute substantively to the conversation, and they explicitly connect their ideas to those put forth by other students. In the example, Leslie, Ben, and Ron all make substantive statements. In lines 1-2, Leslie asks a question to the group about a relevant physics concept. In lines 3-9, Ben provides his thoughts on this concept, making use of a previous example and his understanding of it. In lines 10-12, Ron shares his own reasoning regarding the topic of Leslie's question. Not only do all three students contribute substantively to the discussion in this way; they also make explicit connections across their contributions. In lines 3-9, Ben directly relates his answer to Leslie's preceding question, and in lines 10-12, Ron explicitly connects his answer to Ben's by stating that it is "in other words". The substantive contributions the students make and the way that they connect them indicate that this conversation has an elaborative interaction pattern.

### Disciplinary Content Dimension

The disciplinary content dimension of the Building Understanding towards an Answer Mode is described by abstract physics content. While the students in this context ultimately aim to produce an answer to a question, when they engage in the Building Understanding towards an Answer Mode, they do not focus on this goal directly. Instead, they discuss physics concepts in the abstract or general cases, first establishing an understanding of these before attempting to apply them in the creation of an answer to a particular question. In the example, the students attempt to arrive at an understanding electric potential energy in general and how it relates to electric field lines. In lines 1-2, Leslie begins the episode with a question about the *meaning* of electric potential energy. Her question isn't directly related to determining an answer to the current question of the activity. When Ben responds in lines 3-9, he makes reference to a previous problem the students had completed, but does not attempt to connect it to producing an answer to the current question. In Ron's response in lines 9-10, he provides his understanding of electric potential energy, again, not connecting it to answering the current question in the activity. As the focus of their conversation was not the production of an answer, but instead the meaning and relationships among the concepts involved, their conversation is characterized by abstract physics content.

### Relationships Among Modes

The four Modes of Collaboration defined here can be distinguished from one another along the three dimensions (see Table II). In addition to considering each Mode independently as above, it can be illustrative to consider the similarities and differences among them. One element to consider is the number of participants each Mode tends to have. The Debate Mode is the only Mode which has a specific number of participants – two. Informing, Co-constructing an Answer, and Building Understanding towards an Answer all may have any number of participants (greater than one). The Debate Mode is also unique from the other three Modes in that it is the only one to take place with a dissonant interaction style in the social dimension, while the others have a consonant interaction style.

Co-constructing an Answer and Building Understanding towards an Answer are perhaps the most similar Modes; they both have elaborative interaction patterns in the discursive dimension and consonant interaction styles in the social dimension. Even in the disciplinary content dimension, both Modes ultimately seek to produce answers to the questions in the activity. The crucial distinction between them is the way in which the students go about producing those answers. To illustrate this difference, a continuation of the transcript in Episode D is presented here, in which we see a switch from Building Understanding towards an Answer to Co-constructing an Answer. The fourth member of the group, Donna, who did not participate in Episode D, begins participating here.

|  | **Debate** | **Informing** | **Co-construction of an answer** | **Building understanding towards an answer** |
|---|---|---|---|---|
| **Social** | Dissonant | Consonant | Consonant | Consonant |
| **Discursive** | Argumentation | Consensual | Elaborative | Elaborative |
| **Disciplinary content** | Specific or abstract | Specific | Specific | Abstract |

TABLE II. Summary of the Modes of Collaboration identified in our data

Episode E
13 Leslie: Okay, but then looking too,
14 [gesturing at the image] so this has a large
15 radius away from this section, and it also
16 has like a similar density of lines and like
17 really close to here. So since they the same
18 density of lines and this is a bigger radius,
19 would this be more potential energy?
20 Donna: Well, okay, if you look at like the
21 original lines [gesturing at the image] and
22 the arc length between those, as they get
23 farther out, that shows like… but they
24 added lines in here. I don't know why they
25 did that. But, like that line's kind of added in
26 there. It didn't start from the beginning.
27 Ben: [speaking while Donna continues]
28 Yeah.
29 Donna: This line goes from the beginning.
30 Ron: [speaking while Donna continues]
31 Mmmhmm.
32 Donna: That line goes from the beginning.
33 Leslie: Ohhh. That's confusing, cause then
34 that's not like…
35 Donna: [inaudible] so maybe that's not the
36 same thing.
37 Ben: I don't think that's related to [energy]…
38 is it?
39 Leslie: Related? What do you mean? [looks
40 at Ben]
41 Ben: I was thinking like if I were to like just
42 go this distance [gesturing at the image],
43 like anything… anything in this area would
44 have a greater electric potential energy
45 than like something [gesturing at another
46 area on the image]
47 Donna: Maybe the equipotential lines would
48 have been better.
49 Ben: Yeah.
50 Ron: [inaudible] So where is that? So it
51 would be over the center? Not dead center.
52 [miming being away from the center with
53 his hand]
54 Donna: We could just say the origin of the
55 arrows.
56 Ron: Yeah, the origin of the lines.
57 Leslie: Yeah.
58 [all begin writing on their worksheets]

In contrast to the focus on abstract content in lines 1-12 (discussed in the Building Understanding towards an Answer: Disciplinary Content Dimension section), in lines 13-19, we see Leslie shift the focus of the discussion to the image of the charge distribution from the activity question. Furthermore, her question makes a comparison of the magnitude of the electric potential energy at two points in the image. This is directly related to answering the activity question, which asks the students where the maximum magnitude of electric potential energy in the image occurs. When Donna responds in lines 20-32, she also refers to the image from the activity, and in lines 41-46, Ben discusses relative amounts of electric potential energy at different points in the image, again related directly to answering the activity question. Finally, in lines 50-58, the students explicitly discuss what to write as their final answers on their worksheets, and then do so.

The explicit focus on answering the activity question seen in Episode E indicates that the students are engaging in the Co-constructing an Answer Mode. This is notably different than the Building Understanding towards an Answer seen in Episode D, in which the students instead focused on establishing an understanding of the relevant physics concepts. Episodes D and E also demonstrate the relationship between Building Understanding towards an Answer and the production of an answer to an activity question. In Episode D, the students did not focus directly on producing an answer, but the results of the understanding they had built were later applied to answering an activity question.

Finally, episodes D and E indicate that shifts between Modes might be readily identifiable. Between lines 12 and 13, there was a clear change in the focus of the conversation, which showed a change in the Mode. That the four Modes of Collaboration observed in this context can be defined using the social, discursive, and disciplinary content dimensions, and meaningfully distinguished from one another makes them a framework that could be productively applied to acquire insights not offered by other frameworks, which will be explored in the Discussion.

## DISCUSSION

The Modes of Collaboration can offer a perspective not provided by existing frameworks, but shares elements with and is greatly informed by previous work regarding student group dynamics in physics and other disciplines. In particular, the three dimensions that are used to define the Modes are closely tied to previous work. The social dimension rests on the interaction styles identified by Lumpe and Staver [4], the discursive dimension makes use of the interaction patterns identified by Hogan et al. and argumentation as conceived of by Toulmin [22,33], and the disciplinary content dimension bears similarity to the work on epistemic framing by Irving et al. [13].

The relationship of the Modes of Collaboration framework with the work on epistemic framing by Irving et al. merits particular discussion, as this relationship may be the least clear. The expansive vs narrow axis defined by Irving et al. clearly has a relationship to the abstract and specific content distinguished in the disciplinary content dimension of the Modes; both are related to the degree to which student talk is directly tied to the task on which they are working. While the Modes solely attend to whether or not students are focusing on producing an answer to make a distinction between abstract and specific content, however, that is only one piece of evidence that Irving et al. make use of in their characterization of student discussion on the expansive vs narrow axis. For example, they also consider aspects such as the use of multiple representations to indicate a more expansive framing, and an extended focus on answering an instructor's questions to indicate a narrow framing. In this way, the narrow vs expansive axis is more broadly defined and inclusive of more elements than the abstract and specific physics content distinguished in the disciplinary content dimension. It is also worth noting that the Modes are fundamentally different than the framework for epistemic framing developed by Irving et al. in that they have different goals. The Modes of Collaboration seek to holistically consider and distinguish different types of student interactions in groups based on their behavior, while Irving et al.'s epistemic framing aims to describe how students appear to frame the activity in which they are engaged. To identify a group as engaged in a particular Mode is to describe how they are interacting with each other. This is not necessarily the same as stating that all the students are framing the activity in the same way. One could imagine different students framing a Debate quite differently, for example. Some may see it as a positive manifestation of academic rigor, while others may feel that it is undesirable conflict. In this way, analyzing student behavior through the work on epistemic framing by Irving et al. and using the Modes of Collaboration to identify how students are engaged in group work are separate endeavors.

More generally, the Modes of Collaboration framework is set apart from the previous work on which it builds in that it makes use of the constructs developed by the aforementioned authors simultaneously in the analysis of student interactions, and in a way that distinguishes the social, discursive, and disciplinary content related elements. These three dimensions each carry valuable information on student engagement, and by attending to all three, none of that information is neglected. For example, if one only considered the social and disciplinary contentment dimensions, Informing and Co-construction of an Answer would not be distinguishable, as they are both consonant and focused on specific content. However, to an instructor or researcher, the difference between one student instructing their group on the correct answer, and multiple students working as peers in the development of an answer, can be of great importance.

In addition to retaining the information carried by all three dimensions by attending to them simultaneously, the Modes of Collaboration framework attends to them independently. The framework does not make any a priori assumptions about the relationship between the dimensions, and so analysis along one dimension is not used to inform analysis along the other dimensions. This allows for the possible identification of Modes that may have been neglected had such relationships been assumed. For example, while we did not observe an episode with consonant social interaction and argumentation as its discursive dimension, and such a pairing may seem unlikely, the Modes of Collaboration does not discount its possibility. Attending to the dimensions independently also makes the framework relatively simple to apply, and suitable for multiple types of analyses. One does not need to consider the convolution of the dimensions in identifying Modes, as episodes are categorized along each dimension separately. Once the Modes present have been identified, one could use this data to analyze a snapshot of a single group, compare the Modes present across multiple groups, or consider the Modes a single group engages in over time.

This multi-dimensional approach has benefits when compared to frameworks that focus on a single area. For example, Toulmin analysis focuses on the content and structure of speakers' discourse. Applied in our context, it would be very successful in describing how students constructed their arguments. One thing it would not attend to, however, would be the type of physics the students discussed.

A Debate on specific content and a Debate on abstract content would be treated the same. The multi-dimensional approach used in our framework captures these differences, which can be crucial for understanding the variety of ways that students experience group work.

While the Modes of Collaboration framework is able to provide a unique and productive description of student engagement with group work, it is not an all-encompassing description. Of the video analyzed once moments with off topic conversation or no explicit interaction had been neglected, 68% was identified as being an instance of a Mode. Thus the Modes defined here do not completely describe the possibilities of student interactions. Common interactions not classified as Modes in our data included students confirming how to phrase their answers on their worksheets, and sequences of disconnected statements.

The small number of students and relatively short amount of time analyzed in our data naturally limits the degree to which the Modes might be generalizable. Because the students in our data worked on a single activity, the Modes we observed may have been influenced by that activity. Specifically, the fact that the task required the students to produce answers to turn in on a worksheet may have fostered an environment which favored Modes that focus on the construction of an answer. It is also possible that there is a relationship between the particular physics content in the activity and the Modes in which the students engaged. The highly abstract nature of electric field and potential energy may have prompted students to engage in different Modes than they might if the activity focused on a more familiar or concrete area, such as projectile motion.

Additionally, the classroom in which our data was collected was facilitated with very minimal instructor interaction. It could be that with more active facilitation, or with certain types of facilitation, the Modes observed may vary. In fact, the preliminary analysis of data from a project-based learning environment suggests just this. In the project-based learning environment, facilitators are very active and interact frequently with students, and the tasks that the students work on are complex and open-ended. Observations of this data indicate that these differences in facilitation and activity likely do change the

Modes observed, though further analysis is needed to describe how and why this occurs.

There are other facets of the Modes that may exist, but that we simply did not observe due to our context or the small size of our data. For example, the Co-Construction Mode as observed in this context was specifically the Co-Construction of an Answer. It is possible that other "products" than an answer may be possible in the Co-Construction Mode. Preliminary analysis of the project-based learning environment described earlier suggests that this is the case, as students appear to also Co-Construct calculations, responses to instructor questions, and problem solving strategies. The Debate mode may appear differently in other data, as well. In our context, we only observed Debates between two students, but it is conceivable that a Debate could have more than two students, all arguing unique perspectives, or even that "teams" could form, with multiple students arguing for one perspective and multiple students arguing for another.

Another manifestation of the Modes of Collaboration that may exist, but that we did not observe, is the co-occurrence of Modes. As all of the Modes may have as few as two participants, it is possible that in a group of four, dyads could form and simultaneously engage in different Modes. In our data, when only two students in a group of four participated in a particular Mode, the other two students were not identified as participating in any Mode. They were either quietly observing the two active students, or it was unclear on what their attention was focused. One could imagine, however, that two students in a group could engage in one Mode, while the other two students simultaneously engaged in another. For example, two students could engage in a Debate while the other two students in the group engage in the Co-Construction of an Answer. We did not observe such an episode in our data, but the Modes of Collaboration framework does allow for such an event.

Though it of course cannot describe all of student interactions, the Modes of Collaboration framework shows promise for being a useful tool for instructors. A simple implementation could be identifying which students are participants in episodes of Co-construction of an Answer or Building Understanding towards an Answer and using this as a way to consider how engaged different students are with their work. The Modes may also offer one way to identify if and how students alter their interactions with each other after an intervention by an instructor. Finally, the Modes of Collaboration framework may be useful in comparing how students engage with different facilitation techniques and different activities. As discussed previously, we have made preliminary observations of a different classroom using the framework, and have seen contextual differences and what may be new Modes. It was still possible to analyze these new features using the three dimensions of the Modes of Collaboration, indicating that the framework has the potential to be productively applied across contexts. Doing so across different activity designs or facilitation techniques could provide better insight into the impact that different instructional choices have on student engagement with group work.

As interactive instruction continues to gain popularity, it is essential that instructors and researchers understand how students engage in such activities. The Modes of Collaboration provide a framework allowing insight in this area. The social, discursive, and disciplinary content dimensions on which the framework is based allow it to capture elements of student engagement that other frameworks may not, and also give it the flexibility to be adapted for cross-contextual analysis of different learning environments. This kind of analysis will be crucial for understanding the inherently complex and varied nature of group based learning environments and for ultimately providing students with the best opportunities for learning.